\documentclass[a4paper,11pt]{article}
\usepackage{graphicx, graphics}
\usepackage{latexsym}
\usepackage{cite}
\usepackage{epstopdf}
\usepackage{amssymb}
\author{Parviz Goodarzi \footnote{E-mail: parviz.goodarzi@abru.ac.ir}
\\ {\small\small Department of Physics, School of Sciences}
\\ {\small Ayatollah Boroujerdi University, Boroujerd, Iran}}
\title {Anisotropic inflation in non-minimal kinetic coupling model}

\begin{document}
\maketitle

\begin{abstract}

We study anisotropic inflation in non-minimal derivative coupling model where the scalar field non-minimally coupled to the $U(1)$ gauge
fields and derivative of the scalar field non-minimally coupled to the Einstein tensor.
Within the framework we find power-law anisotropic solutions in this model when both the inflaton potential and the gauge kinetic function are power-law type in the high friction regime.
We show the ratio of anisotropy to the expansion rate is nearly constant, small and proportional to the slow-roll parameters of the theory.
As a demonstration, we consider numerically calculation of the model to show that the behavior of anisotropy by changing the parameters of the model for quadratic inflationary potential.
There is anisotropic attractor solution for a wide range of values of the model parameters.
We show both numerically and analytically that there are two phases of inflation, similar to those an anisotropic inflation in minimal coupling model, isotropic and anisotropic phase.
We can change the number of e-folds corresponding to each phase of slow roll inflation by changing the gauge coupling constant or non-minimal derivative coupling constant.
There are the best agreement with the numerically solutions and analytically solutions in this investigation.

 \end{abstract}
\noindent\rule{11cm}{0.4pt}

\tableofcontents
\noindent\rule{11cm}{0.4pt}

\section{Introduction}

  The recent data from cosmic microwave background and the examination of the large scale structures shows that the Universe is homogeneous and isotropic at large scales but it has anisotropy and non homogeneity at tiny scales \cite{Planck,Planck1,WMAP,WMAP1,WMAP2}.
Inflation is one of the most successful paradigms that explains not only homogeneity and isotropy at large scale
but also, it has the best answers for anisotropy in cosmic microwave background and large scale structures formation in Universe \cite{Guth,inflaton1,Weinberg,Mukhanov,Liddle}.
In recent decades the most models which have been proposed for inflation where based on minimally or
non-minimally coupled canonical scalar field to gravity.
A scalar field that generates inflation slow-rolling down to minimum of the potential
where the potential energy of scalar field similar to energy density of cosmological constant causes
the accelerated expansion of the Universe exhibits exact de Sitter nature \cite{Lyth,Kolb}.
In addition, this simple scenario predicts that primordial inhomogeneities of the universe are almost isotropic, adiabatic, approximately scale invariant and nearly Gaussian \cite{Maldacena}.

On the other hand, "cosmic no-hair conjecture" for a model with a positive cosmological constant says that any anisotropy, except for quantum fluctuation as a seed of large scale structures of the universe, will be exponentially damped by the dynamics of the system during inflation in a few Hubble time scale \cite{Wald}. In other words, inflation washed away the anisotropies, regardless of the inflationary models \cite{Barrow,Sheikh-Jabbari1}.

However, with a more detailed examination of the WMAP data the authors of references \cite{Groeneboom,Hanson,Groeneboom2,Carroll} concluded that there is an anomaly related to the broken rotational invariance of the CMB perturbations. So that the simplest single scalar field inflation could not explain deviation from temporal isotropy and Gaussian nature of primordial perturbations in cosmic microwave background.

Nevertheless, motivated by these finding, several theoretical propositions came up to explain anisotropy generated during the inflation.
In particular, multifield models of inflation was noticed in extension of the model building of inflation, for example, in fundamental theories such as supergravity and string theory.

The inflationary models with the vector fields and the two form fields in the context of supergravity have been suggested that anisotropic hair can remain to the end of inflation which result in is a counter example to the cosmic no hair conjecture \cite{Ohashi,Tsujikawa,Soda}.
Therefore vector fields have a substantial role in statement of anisotropic inflationary cosmology.
In these models we have seen a coupling between the scalar field and derivative of $U(1)$ gauge field
in the form of $f(\varphi)^2F_{\mu\nu}F^{\mu\nu}$. However, to have a growing but sub-dominant
anisotropy, the coupling function must be suitably chosen which causes stable anisotropic inflation \cite{Soda,Soda1}.
The existence magnetic field during inflation generically breaks the rotational invariance that will make preferable direction of space-time \cite{Soda2,Sheikh-Jabbari}. There exist an inflationary attractor solutions along which the anisotropy shear is nearly constant \cite{Soda3}.

Therefore these are the enough motivations to propose the inflationary models that generate statistical anisotropy naturally.
In the ref. \cite{Sheikh-Jabbari} it has been shown that the non Abelian gauge fields $SU(2)$ with the Bianchi type-I background metric
have a anisotropic attracting behavior for generic initial values.
Non-minimal anisotropic inflation in the general relativity and teleparallel framework of gravity have been considered in \cite{Abedi,Chen} respectively.
Anisotropic inflation in Gauss-Bonnet gravity, in Brans-Dicke gravity and with the coupling between derivative of scalar field and $U(1)$ gauge fields have been investigated extensively in the literature \cite{Sayantani,Tirandari,Holland}.
In ref. \cite{Firouzjahi}, the effect of charged scalar field in anisotropic inflation has been considered.
In ref. \cite{Ghalee}, the coupling between the Einstein tensor and the derivative of scalar fields in Bianchi type-I background has been
investigated, but in that work, the author ignore the effects of $U(1)$ gauge fields.

Non-minimal kinetic coupling $G^{\mu \nu}\partial_\mu\varphi \partial_{\nu} \varphi$ is one of the operators of Horndeski’s scalar-tensor theory where primordially introduced in \cite{Horndeski, Charmousis} and used in references \cite{Germani1,Germani2} to Higgs inflation. In this coupling the inflaton field evolves more slowly relative to the case of standard inflation due to a gravitationally enhanced friction which its capacity to explain the Higgs inflation. Moreover this coupling is also safe of quantum corrections and unitary violation problem, without introducing new degrees of freedom \cite{Germani3,Germani4,Tsujikawa2,Sadjadi,Sadjadi1}.

An emphasize feature of the non-minimal derivative coupling with the Einstein tensor is that the mechanism of the gravitationally enhanced friction during inflation, by which even steep potentials with theoretically natural model parameters can drive cosmic acceleration \cite{Germani4, Tsujikawa2}.
Thus it is well motivated to propose the anisotropic inflation in context of non-minimal derivative coupling model.

Hence, in this paper, we examine anisotropic inflation in the presence of the $U(1)$ gauge fields in context of non-minimal derivative coupling model and the effect of "gravitationally enhanced friction" on the evolution of the scalar field with the anisotropic hair has been considered.
In order to have non-vanishing anisotropy, the form of coupling function $f(\varphi)$ should include an exponential function of the inflaton potential and the coupling constant.
We consider numerical analysis for the quadratic potential and we have seen that there are two phases of inflation, isotropic and anisotropic phase. We have seen that this mechanism enhance the number of e-folds during isotropic phase of inflation.

This paper is organized as follows.
In section 2 we review non-minimal derivative coupling in Bianchi type-I geometry in the presence of $U(1)$ gauge fields, and we derive the basic equations of motion for scaler field on the anisotropic cosmological background.
In section 3 we obtain power-law solutions for those equations in a high friction regime with the power-law coupling function and scaler field potential.
In section 4 we consider slow-roll anisotropic inflation in this model and we obtain the anisotropy shear as a function of slow-roll parameters and coupling function as a exponential function of potential.
In section 5 we depicted equations of motion numerically for quadratic potential and it show that, there is an attractor solutions for any different values of the parameters of model and we compare the numerical results with the analytical results.
Conclusion and brief discussions are given in the final section.

We use natural units $\hbar=c=1$ though the paper.

\section{The models}

In this section, we will introduce the action and we obtain the equations of motion for our model.
We add an abelian gauge field with $\varphi$-dependent coupling function $f(\varphi)^2$ to Gravitational Enhanced Friction (GEF)
theory which is described by an action \cite{Germani1}. Therefore the action under consideration is written as
\begin{equation}\label{1}
S=\int \bigg({M_P^2\over 2}R-{1\over 2}\Delta^{\mu \nu}\partial_\mu
\varphi \partial_{\nu} \varphi- V(\varphi)-{1\over 4}f(\varphi)^2F_{\mu\nu}F^{\mu\nu}\bigg) \sqrt{-g}d^4x.
\end{equation}

Where $\Delta^{\mu \nu}=g^{\mu \nu}+(1/M^2)G^{\mu \nu}$, $G^{\mu \nu}=R^{\mu \nu}-{1\over 2}Rg^{\mu \nu}$ is the Einstein
tensor, $R$ is the Ricci scalar, $M$ is a coupling constant with the dimension of mass, $M_P=1/\sqrt{8\pi G}=2.4\times 10^{18}GeV$ is the reduced Planck
mass, $F_{\mu\nu}=\partial_{\mu}A_{\nu}-\partial_{\nu}A_{\mu}$ is the field strength tensor of the $U(1)$ gauge field $A_\mu$, and $f(\varphi)$ describes the coupling function of the $U(1)$ gauge fields and scalar field $\varphi$.
In this work, we will refer to the field strength $F_{\mu\nu}$ as the electromagnetic field.
Also we assume that the background is homogeneous and there is a nonzero homogeneous electric field.
In order to have gauge invariance, we choose the Coulomb-radiation gauge $A_0=\partial_iA^i=0$ and
without loss of generality, we choose the x-axis as the direction of vector field.
So that we take the homogeneous vector field of the form $A_{\mu}=(0,v(t),0,0)$ where $v(t)$ is arbitrary function of cosmic time.
By this choose we have $F_{ij}=F_{\mu 2}=F_{\mu 3}=0$ and $F_{01}=E_x=-\dot{v}\neq 0$ where $E_x$ is the electric field in the x direction.
In other words, there is a nonzero homogeneous electric field in the x-axis direction.
This field configuration holds the plane symmetry in the plan perpendicular to the vector field.
The background space time is Bianchi type-I where the anisotropic line element can be written in the form of
\begin{eqnarray}\label{2}
ds^2=-\mathcal{N}^2dt^2+\nonumber \\
e^{2\alpha(t)}\bigg[e^{-4\beta(t)}dx^2&+
e^{2\beta(t)}\Big(e^{2\sqrt{3}\beta_{-}(t)}dy^2+
e^{-2\sqrt{3}\beta_{-}(t)}dz^2\Big)\bigg].
\end{eqnarray}
Here $e^{\alpha(t)}$ is the isotropic scale factor where $\alpha(t)$ measures the number of e-foldings, $e^{\beta(t)}$ and $e^{\beta_{-}(t)}$ characterize the anisotropy
and $\mathcal{N}$ is the laps function introduced for calculation of Hamiltonian constraint.
Therefore, the metric becomes
\begin{eqnarray}\label{3}
g_{\mu\nu}=diag\bigg[-\mathcal{N}^2,e^{2\alpha(t)-4\beta(t)},e^{2\alpha(t)+2\beta(t)+2\sqrt{3}\beta_{-}(t)},&& \nonumber \\
e^{2\alpha(t)+2\beta(t)-2\sqrt{3}\beta_{-}(t)}\bigg].
\end{eqnarray}
So determinant of the metric is $g=-\mathcal{N}^2e^{6\alpha(t)}$ and time-time component of the Einstein tensor becomes
\begin{equation}\label{4}
G_{tt}=3\Big(\dot{\alpha}^2-\dot{\beta}^2-\dot{\beta_{-}}^2\Big).
\end{equation}
Where the overdot denotes differentiation with respect to the cosmic time $t$. By substituting the metric (\ref{3}) into the action (\ref{1}) we have
\begin{eqnarray}\label{5}
S&=&\int e^{3\alpha(t)}\bigg[
{3M_P^2\over\mathcal{N}^2}\Big(2\mathcal{N}\dot{\alpha}^2+\mathcal{N}\dot{\beta}^2+\mathcal{N}\dot{\beta_{-}}^2+\mathcal{N}\ddot{\alpha}-
\dot{\mathcal{N}}\dot{\alpha}\Big) \nonumber\\
&&+{{\dot{\varphi}}^2\over2\mathcal{N}}\bigg(1+{3\over M^2{\mathcal{N}}^2}\Big({\dot{\alpha}}^2-{\dot{\beta}}^2-{\dot{\beta_{-}}}^2\Big)\bigg)
-\mathcal{N}V(\varphi)\nonumber\\
&&+{1\over2\mathcal{N}}f(\varphi)^2\dot{v}^2e^{-2\alpha+4\beta}\bigg]d^4x.
\end{eqnarray}
Using integration by parts we can rewrite the action (\ref{5}) as
\begin{eqnarray}\label{6}
S&=&\int{e^{3\alpha(t)}\over\mathcal{N}}\bigg[3M_P^2\Big(-\dot{\alpha}^2+\dot{\beta}^2+\dot{\beta_{-}}^2\Big)  \nonumber\\
&&+{{\dot{\varphi}}^2\over2}\bigg(1+{3\over M^2{\mathcal{N}}^2}\Big({\dot{\alpha}}^2-{\dot{\beta}}^2-{\dot{\beta_{-}}}^2\Big)\bigg) \nonumber \\
&&-{\mathcal{N}}^2V(\varphi)+{1\over2}f(\varphi)^2\dot{v}^2e^{-2\alpha+4\beta}\bigg]d^4x.
\end{eqnarray}
The variation of action (\ref{6}) with respect to $\beta_-$ yields
\begin{equation}\label{7}
\dot{\beta_{-}}(t)=C{e^{-3\alpha(t)}\over\Big(M_p^2-{{\dot{\varphi}}^2\over 2M^2}\Big)},
\end{equation}
where $C$ is the constant of integration. This relation demonstrates that an anisotropy in $y-z$ plane
rapidly decay with the expansion of the universe during inflation. Then, we ignore the effects of this anisotropy by choosing $C=0$.
Therefore, the line element becomes
\begin{equation}\label{8}
ds^2=-\mathcal{N}^2dt^2+e^{2\alpha(t)}\Big[e^{-4\beta(t)}dx^2+e^{2\beta(t)}(dy^2+dz^2)\Big].
\end{equation}
Now, through the variation of this action with respect to $v$ we can obtain $\dot{v}$ in gauge field as
\begin{equation}\label{9}
\dot{v}=P_Af(\varphi)^{-2}e^{-\alpha(t)-4\beta(t)},
\end{equation}
where $P_A$ is the constant of integration. By taking the variation of the action (\ref{6}) with respect to $\mathcal{N},\alpha, \beta$ and $\varphi$, then putting the lapse function equals to one $\mathcal{N}=1$ and substituting $\dot{v}$ from relation (\ref{9}),
we obtain the following four field equations
\begin{eqnarray}
\label{10}
\dot{\alpha}^2-\dot{\beta}^2-{1\over3M_p^2}\bigg[{{\dot{\varphi}}^2\over2}\Big(1+{9\over M^2}({\dot{\alpha}}^2-
{\dot{\beta}}^2)\Big)+V(\varphi)&& \nonumber \\
+{P_A^2\over2}f(\varphi)^{-2}e^{-4\alpha-4\beta}\bigg]&=&0,\\
\label{11}
\Big(2\ddot{\alpha}+3\dot{\alpha}^2+3\dot{\beta}^2\Big)\Big[{{\dot{\varphi}}^2\over2M^2}-M_p^2\Big]
+{2\over M^2}\dot{\alpha}\dot{\varphi}\ddot{\varphi}-{{\dot{\varphi}}^2\over2}+V(\varphi) \nonumber \\
-{P_A^2\over6}f(\varphi)^{-2}e^{-4\alpha-4\beta}&=&0,\\
\label{12}
\Big(\ddot{\beta}+3\dot{\beta}\dot{\alpha}\Big)\Big[{{\dot{\varphi}}^2\over 2M^2}-M_p^2\Big]+\dot{\beta}{\dot{\varphi}\ddot{\varphi}\over M^2}+{P_A^2\over3}f(\varphi)^{-2}e^{-4\alpha-4\beta}&=&0,\\
\label{13}
\Big(\ddot{\varphi}+3\dot{\alpha}\dot{\varphi}\Big)\Big[1+{3\over M^2}({\dot{\alpha}}^2-{\dot{\beta}}^2)\Big]
+{6\over M^2}\dot{\varphi}\Big(\dot{\alpha}\ddot{\alpha}-\dot{\beta}\ddot{\beta}\Big)+V'(\varphi) \nonumber\\
-P_A^2f(\varphi)^{-3}f'(\varphi)e^{-4\alpha-4\beta}&=&0.
\end{eqnarray}

Where prime represents the derivative with respect to the scalar field $\varphi$. The relation (\ref{10}) is the Hamiltonian constraint.
Therefore the energy density of vector field is
\begin{equation}\label{14}
\rho_v={1\over2}P_A^2f(\varphi)^{-2}e^{-4\alpha-4\beta}.
\end{equation}
Indeed, if we take $P_A=\dot{\beta}=0$, in the equations (\ref{10}-\ref{14}) these equations reduces to the equations of motion in non-minimal derivative coupling models \cite{Germani1}.
Also if we take $1/M^2=0$ equations of motion for standard anisotropic inflation in general relativity is obtained \cite{Soda}.
These equations are not independent equations, so that one of these equations is derived from other equations.

\section{Power-law anisotropic solutions in high friction regime}

\begin{figure}[t]
\begin{center}
 \scalebox{0.5}{\includegraphics{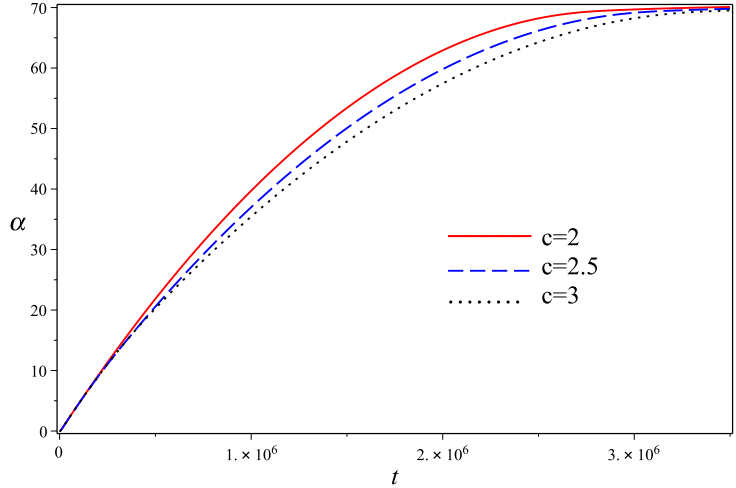}}
  \end{center}
  \caption{The number of e-folds $\alpha$ in terms of cosmic time $M_pt$ in quadratic inflationary potential for coupling constant $1/M^2=10^4$ and different values of $c$.}
   \label{fig1}
\end{figure}
\begin{figure}[t]
\begin{center}
    \scalebox{0.5}{\includegraphics{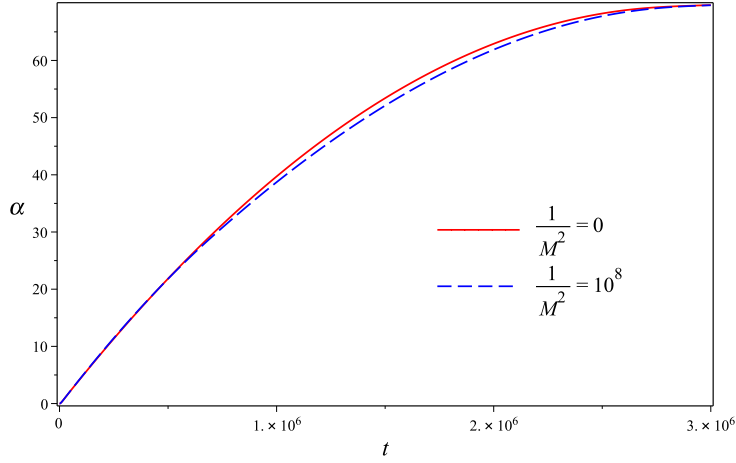}}
\end{center}
      \caption{The number of e-folds $\alpha$ in terms of cosmic time $M_pt$ in quadratic inflationary potential for $c=2$ and different values of coupling constant $M$.}
\label{fig2}
\end{figure}

In this section, we shall consider the power-law solutions for anisotropic inflation in the context of non-minimal derivative coupling model.
We assume that inflaton potential and gauge kinetic function are power-law
\begin{equation}\label{15}
V(\varphi)=\lambda \varphi^n,\;\;\;\;\;\;\; f(\varphi)=f_0 \varphi^m,
\end{equation}
where $\lambda,n,f_0,m$ are constant parameters of the theory.
Let us first look for the power-law solutions by assuming
\begin{equation}\label{16}
\alpha=\zeta\log{M_pt},\;\;\;\; \beta=\eta\log{M_pt}, \;\;\;\;\varphi=\kappa t.
\end{equation}
By substituting of these relations into the Hamiltonian constraint (\ref{10}) in the high friction regime, ${{\dot{\alpha}}^2/ M^2\gg1}$ we have
\begin{eqnarray}\label{17}
\Big(\zeta^2-\eta^2\Big)\Big[1-{3\kappa^2\over2M^2M_p^2}\Big]t^{(-2)}&=&\Bigg({\lambda\kappa^n\over3M_p^2}\Bigg)t^n+ \nonumber \\
&&\Bigg({P_A^2f_0^2\kappa^{-2m}\over 6M_p^{2+4\zeta+4\eta}}\Bigg)t^{(-2m-4\zeta-4\eta)}.
\end{eqnarray}

Therefore, to have the same power of time for each term in two sides of this equation, we get two conditions as;
\begin{equation}\label{18}
n=-2,\;\;\;\;\;\;\;\;\; m+2\zeta+2\eta=1,
\end{equation}
also, to have the same amplitude in the two sides of this equation, we need
\begin{equation}\label{19}
\Big(\zeta^2-\eta^2\Big)\Big[1-{3\kappa^2\over2M^2M_p^2}\Big]={\lambda\kappa^n\over3M_p^2}
+{P_A^2f_0^2\kappa^{-2m}\over 6M_p^{2+4\zeta+4\eta}},
\end{equation}
where we have defined new variables as;
\begin{equation}\label{20}
x={\kappa^2\over M^2M_p^2},\;\;\;\;\;u={\lambda\kappa^n\over M_p^2}, \;\;\;\;\;\; w={P_A^2f_0^2\kappa^{-2m}\over M_p^{2+4\zeta+4\eta}}.
\end{equation}
From the equation (\ref{19}) we can obtain
\begin{equation}\label{21}
\Big(\zeta^2-\eta^2\Big)\Big(1-{3\over2}x\Big)={u\over3}+{w\over6}.
\end{equation}
Similarly we can rewrite equations (\ref{11}-\ref{13}) in the form of;
\begin{eqnarray}
\label{22}
\Big(-2\zeta+3\zeta^2+3\eta^2\Big)\Big(1-{1\over2}x\Big)&=&u+{w\over6},\\
\label{23}
\Big(-\eta+3\zeta\eta\Big)\Big(1-{1\over2}x\Big)&=&{w\over3},\\
\label{24}
3\Big(3\zeta-4\Big)\Big(\zeta^2-\eta^2\Big)x&=&2u+mw.
\end{eqnarray}
From relation (\ref{18}) we have $\eta={1/2}-{m/2}-\zeta$. Hence by substituting this relation into the equation (\ref{23}) we can solve $w$ as
\begin{equation}\label{25}
w={3\over2}\Big(2-x\Big)\Big(3\zeta-1\Big)\Big({1\over2}-{m\over2}-\zeta\Big).
\end{equation}
Substituting this result into equation (\ref{21}), we obtain $u$ as follows
\begin{eqnarray}\label{26}
u=3\bigg[\zeta^2-\Big({1\over2}-{m\over2}-\zeta\Big)^2\bigg]\Big(1-{3x\over2}\Big) \nonumber \\
-{3\over4}\Big(2-x\Big)\Big(3\zeta-1\Big)\Big({1\over2}-{m\over2}-\zeta\Big).
\end{eqnarray}
Putting equations (\ref{25}) and (\ref{26}) into equation (\ref{22}) we have
\begin{equation}\label{26.1}
x={-1+4m+6\zeta-3m^2-6m\zeta\over -2+5m+9\zeta-3m^2-9m\zeta}.
\end{equation}
Finally, using equations (\ref{26.1}) and (\ref{24}) we get
\begin{eqnarray}\label{27}
&&{\Big(3\zeta-4\Big)\over\Big(3m+9\zeta-2\Big)}\times\nonumber\\
&&\Big(m-1+4\zeta\Big)\Big[1-4m-6\zeta+3m^2+6m\zeta\Big]=\nonumber\\
&&{\Big(m-1+4\zeta\Big)\over\Big(3m+9\zeta-2\Big)}\times \nonumber\\
&&\Big[-2+8m-6m^2+15\zeta-24m\zeta-18\zeta^2+9m^2\zeta+18m\zeta^2\Big].
\end{eqnarray}
The solutions of the aforementioned equation for $\zeta$ are
\begin{equation}\label{28}
\zeta_I={1\over4}\Big(1-m\Big), \;\;\;\;\;\ \zeta_{II}={1\over2}\Big({1\over3}-m\Big),
\end{equation}
that their corresponding $\eta$'s using equation (\ref{18}) are obtained as follows
\begin{equation}\label{29}
\eta_I={1\over4}(1-m),  \;\;\;\;\;\;\;\;\ \eta_{II}={1\over3}.
\end{equation}
In order to avoid the singularity of relation (\ref{27}) there is a condition for $\zeta$ as follows
\begin{equation}\label{30}
\zeta\neq{1\over3}\Big({2\over3}-m\Big).
\end{equation}
From the expression for scale factor $a=e^{\alpha(t)}$, the Hubble parameter becomes $H={\dot{a}/a}=\dot{\alpha}$. To look for inflationary solutions we need to define the slow-roll parameter $\varepsilon$ in terms of the Hubble parameter $H$, as
\begin{equation}\label{31}
\varepsilon=-{\dot{H}\over H^2}=-{\ddot{\alpha}\over\dot{\alpha}^2}={1\over \zeta}.
\end{equation}
In order to have slow-roll inflation, we need $\varepsilon\ll1$. Hence, $\zeta$ must be much larger then $1$.
From solution (\ref{28}), it turns out that this requirement can be achieved by assuming
\begin{equation}\label{32}
m\ll-1.
\end{equation}

Now, the anisotropy is specified by
\begin{equation}\label{33}
{\Sigma\over H}\equiv{\dot{\beta}\over\dot{\alpha}}={\eta\over \zeta}.
\end{equation}
Let us now discuss two cases of interest.
First, $\zeta=\zeta_I, \eta=\eta_I$,
 as we can see in this case $\Sigma/H=\eta_I/\zeta_I\approx1$.
 This indicates that during the expansion of the universe anisotropy shear is nearly constant and very large.
 Therefore, this solution is not our desired solution.

Second, $\zeta=\zeta_{II}, \eta=\eta_{II}$,
 in this case we have a remarkable result for the shear, as follows
\begin{equation}\label{33.1}
{\Sigma\over H}={\eta_{II}\over\zeta_{II}}\approx{2\over1-3m}\approx{1\over3}\varepsilon.
\end{equation}
We have seen in this case the anisotropy is proportional to slow-roll parameter. Hence, the anisotropic shear is small from the condition (\ref{32}).

The ratio of the energy density of the vector field to that of inflaton field during this periods is nearly constant
\begin{equation}\label{33.2}
{\rho_v\over \rho_\varphi}\approx{P_A^2f(\varphi)^{-2}e^{(-4\alpha-4\beta)}\over 2V(\varphi)}\approx{P_A^2f_0^{-2}M_p^{(-4\zeta-4\eta)}\over 2\lambda \kappa^{(2m-2)}}.
\end{equation}

Since we are only interested in solutions where anisotropy is small and proportional to the slow-roll parameter, this case is our desired solution.

\begin{figure}[t]
\begin{center}
  \scalebox{0.3}{\includegraphics{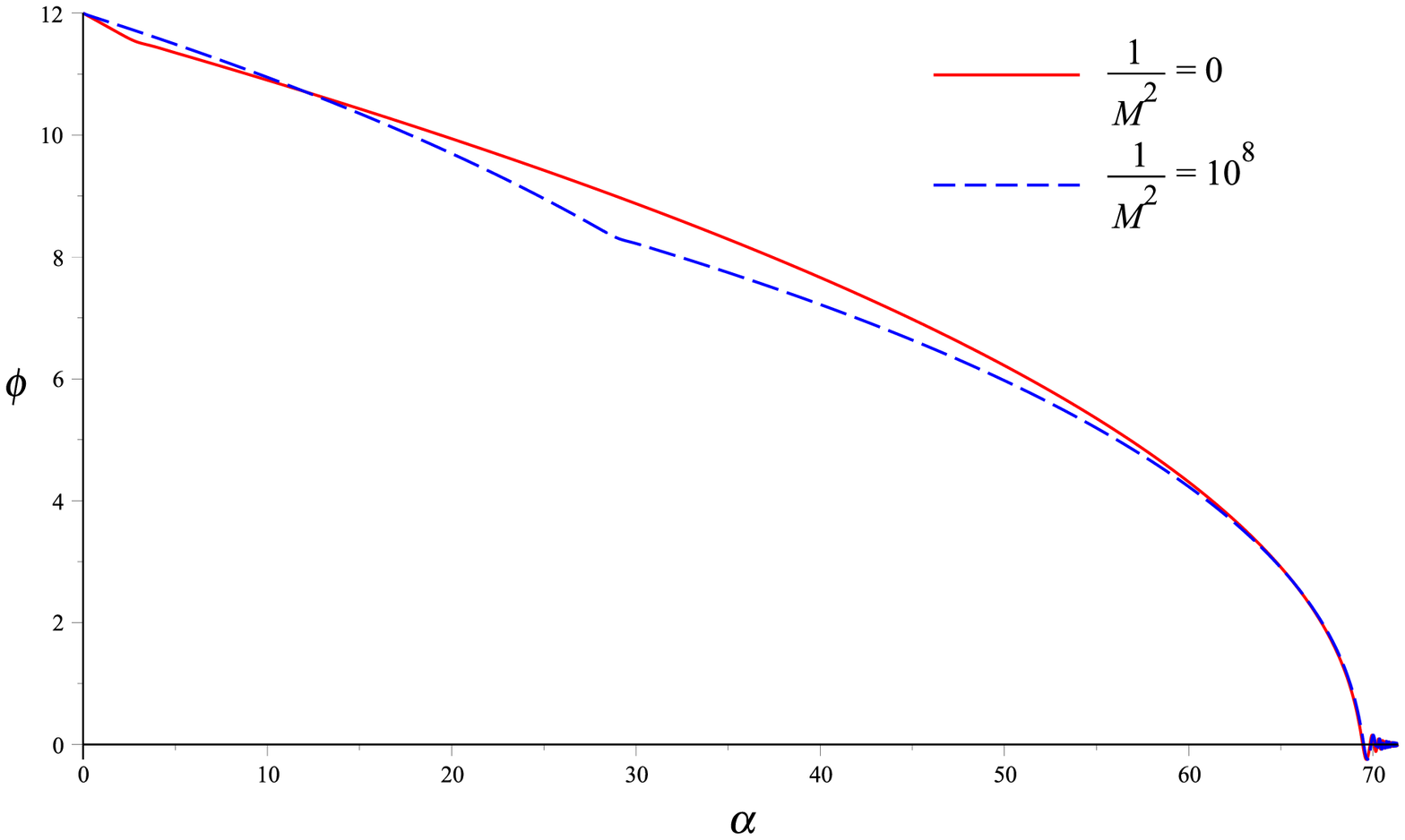}}
  \scalebox{0.34}{\includegraphics{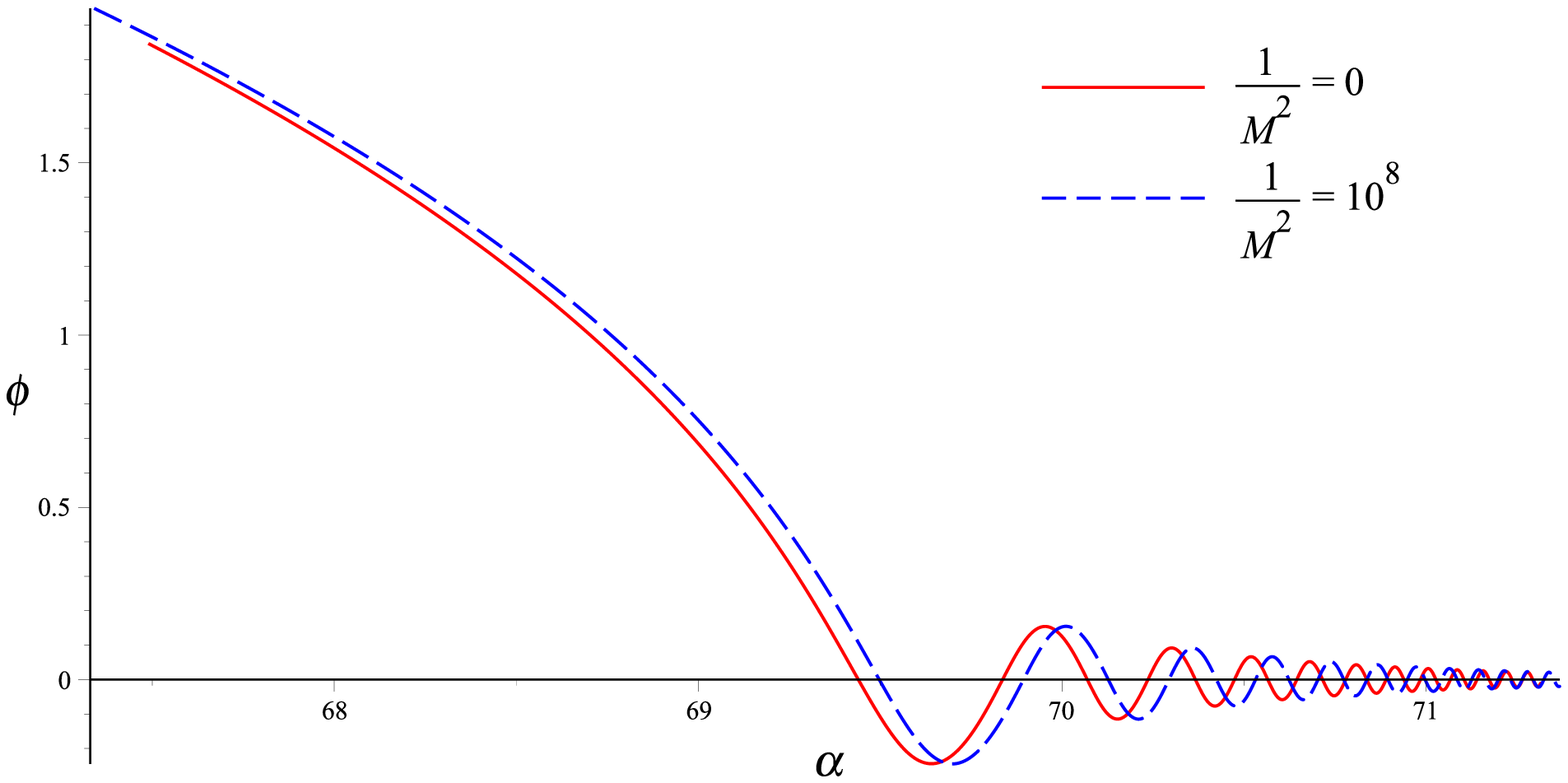}}
\end{center}
   \caption{${\varphi}$ in terms of the number of e-folds $\alpha$, in quadratic inflationary potential for $c=2$ and different values of coupling constant $M$.}
\label{fig3}
\end{figure}

\section{Anisotropic slow-roll inflation}

\begin{figure}[t]
\begin{center}
  \scalebox{0.3}{\includegraphics{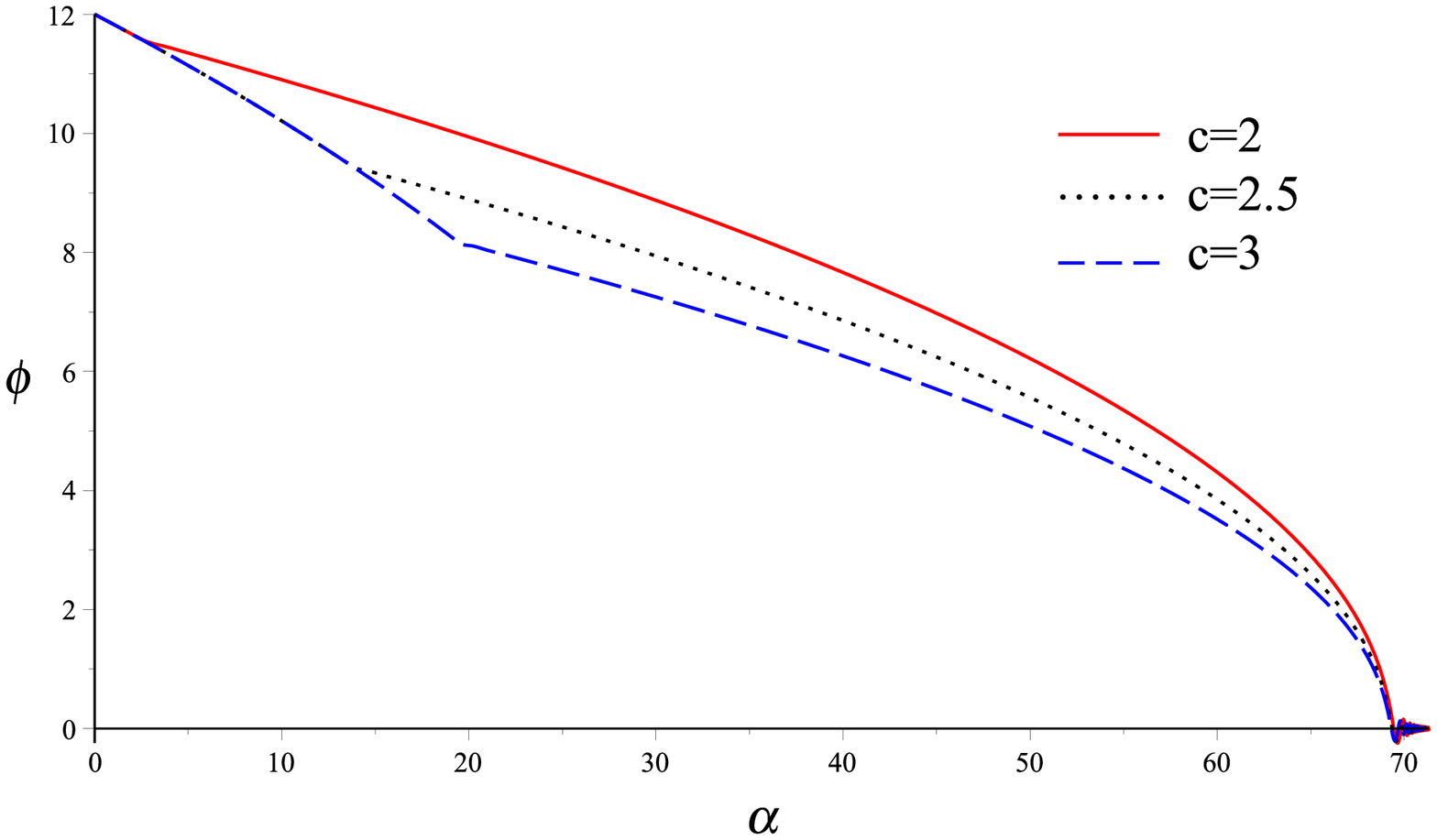}}
  \scalebox{0.3}{\includegraphics{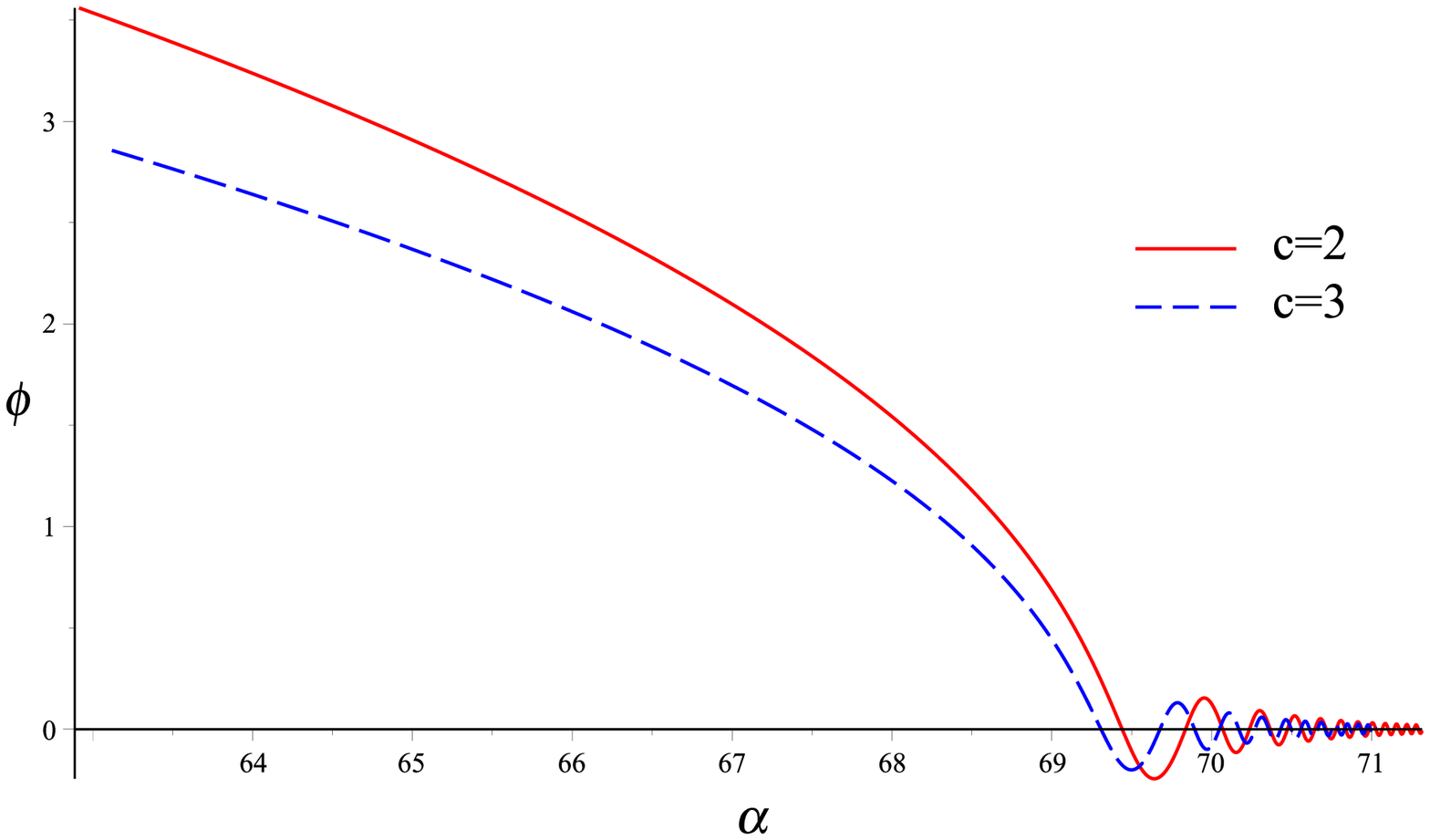}}
\end{center}
   \caption{${\varphi}$ in terms of the number of e-folds $\alpha$, in quadratic inflationary potential for coupling constant $1/M^2=10^4$ and different values of $c$.}
\label{fig4}
\end{figure}

In this section we shall study the generality of slow-roll anisotropic inflation in presence of non-minimal derivative coupling term.
We want to find the appropriate gauge coupling function that generates the anisotropy proportional to the slow-roll parameter.
Let us denote the anisotropic shear $\Sigma\equiv\dot{\beta}$ and the Hubble parameter $H\equiv\dot{\alpha}$ for simplicity.
Now we rewrite the equation (\ref{12}) in terms of $\Sigma$ and $H$ as
\begin{equation}\label{34}
\Big(\dot{\Sigma}+3H\Sigma\Big)\Big[1-{{\dot{\varphi}}^2\over 2M^2M_p^2}\Big]-{\dot{\varphi}\ddot{\varphi}\over M^2M_p^2}\Sigma-{2\over3M_p^2}\rho_v=0,
\end{equation}
where $\rho_v$ is the energy density of vector field defined in equation (\ref{14}).
For slow roll inflation, it is required that
\begin{equation}\label{35}
\dot{\Sigma}\ll3H\Sigma, \;\;\;\;\;\;\;  \ddot{\varphi}\ll 3H\dot{\varphi}.
\end{equation}
Applying the slow-roll conditions to the equation (\ref{34}), we obtain
\begin{equation}\label{36}
3H\Sigma\Big[1-{{\dot{\varphi}}^2\over 2M^2M_p^2}\Big]\approx{2\over3M_p^2}\rho_v.
\end{equation}
Let us work in the small anisotropic limit, so that $\Sigma\ll H$ and $\rho_v\ll\rho_{\varphi}$.
The potential of inflaton dominates during inflation, i.e.
\begin{equation}\label{37}
{{\dot{\varphi}}^2\over2}\Big[1+{9\over M^2}\Big({\dot{\alpha}}^2-{\dot{\beta}}^2\Big)\Big] \ll V(\varphi).
\end{equation}
Under these conditions, the hamiltonian constraint equation (\ref{10}) yields
\begin{equation}\label{38}
H^2\approx{V(\varphi)\over3M_p^2}.
\end{equation}
Using equations (\ref{36}) and (\ref{38}) we find the ratio of the shear to the Hubble parameter
\begin{equation}\label{39}
{\Sigma\over H}\approx{2\rho_v\over3V(\varphi)}\bigg(1-{{\dot{\varphi}}^2\over 2M^2M_p^2}\bigg)^{-1}.
\end{equation}
Now, by applying slow-roll conditions to the equation (\ref{13}), we obtain
\begin{equation}\label{40}
3H\dot{\varphi}\bigg(1+{3H^2\over M^2}\bigg)+V'(\varphi)\approx 0.
\end{equation}
From relations (\ref{38}) and (\ref{40}) we have
\begin{equation}\label{41}
\dot{\varphi}^2\approx {M_p^2V'(\varphi)^2\over 3V(\varphi)}\bigg(1+{V(\varphi)\over M^2M_p^2}\bigg)^{-2}.
\end{equation}
Again, using equations (\ref{39}) and (\ref{41}) we can calculate the ratio of the anisotropic shear to the Hubble parameter as a function of inflaton potential and energy density of vector field as
\begin{equation}\label{42}
{\Sigma\over H}\approx{2\rho_v\over3V(\varphi)}\Bigg[1-{V'(\varphi)^2\over 6M^2V(\varphi)}\bigg(1+{V(\varphi)\over M^2M_p^2}\bigg)^{-2}\Bigg]^{-1}.
\end{equation}
Using equations (\ref{38}) and (\ref{40}) we have
\begin{equation}\label{43}
{\dot{\alpha}\over \dot{\varphi}}={d{\alpha}\over d{\varphi}}\approx-{1\over M_p^2}{V(\varphi)\over V'(\varphi)}\bigg(1+{V(\varphi)\over M^2M_p^2}\bigg).
\end{equation}

Therefore by solving of this differential equation we can obtain $\alpha$ as a function of scalar field $\varphi$ as

\begin{equation}\label{44}
\alpha=-{1\over M_p^2}\int{V(\varphi)\over V'(\varphi)}\bigg(1+{V(\varphi)\over M^2M_p^2}\bigg)d\varphi.
\end{equation}

The challenge of this model is to choose of gauge coupling function $f(\varphi)$, such that energy density of vector field $\rho_v$ ,from relation (\ref{14}), be approximately constant during the slow-roll inflation. We saw that a choice guaranteed to work is the simple coupling function $f(\varphi)=e^{-2c\alpha}$ \cite{Soda,Soda1}. Substituting $\alpha$ from equation (\ref{44}) into this coupling function, we have
\begin{equation}\label{45}
f(\varphi)=\exp{\bigg[{2c\over M_p^2}\int{V(\varphi)\over V'(\varphi)}\bigg(1+{V(\varphi)\over M^2M_p^2}\bigg)d\varphi\bigg]},
\end{equation}
where $c$ is the constant parameter near 1. Parameter $c$ is obtained by differentiating the coupling function $f(\varphi)$ with respect to $\varphi$ as follows
\begin{equation}\label{46}
c={M_p^2\over2}{f^{'}(\varphi)V^{'}(\varphi)\over f(\varphi)V(\varphi)}\bigg(1+{V(\varphi)\over M^2M_p^2}\bigg)^{-1}.
\end{equation}
\begin{figure}[t]
\begin{center}
  \scalebox{0.6}{\includegraphics{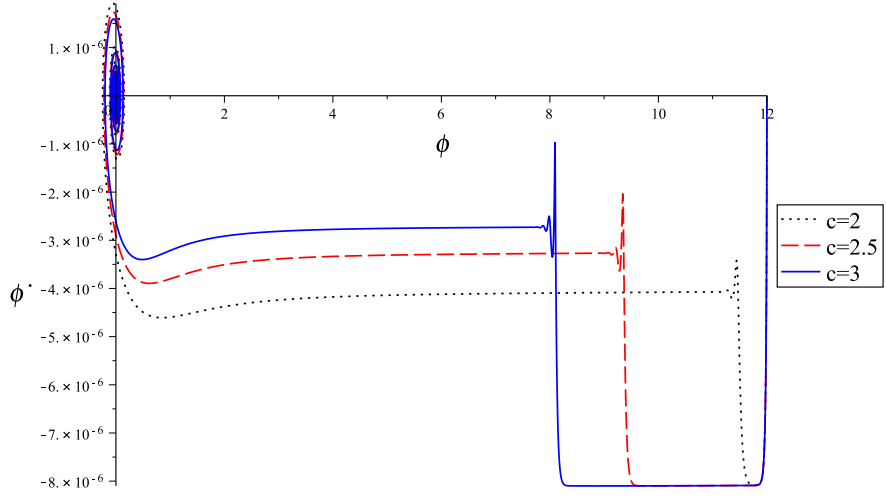}}
\end{center}
  \caption{The phase flow for $\phi$, in quadratic inflationary potential, for coupling constant $1/M^2=10^4$ and different values of $c$.}
\label{fig5}
\end{figure}
\begin{figure}[t]
\begin{center}
  \scalebox{0.6}{\includegraphics{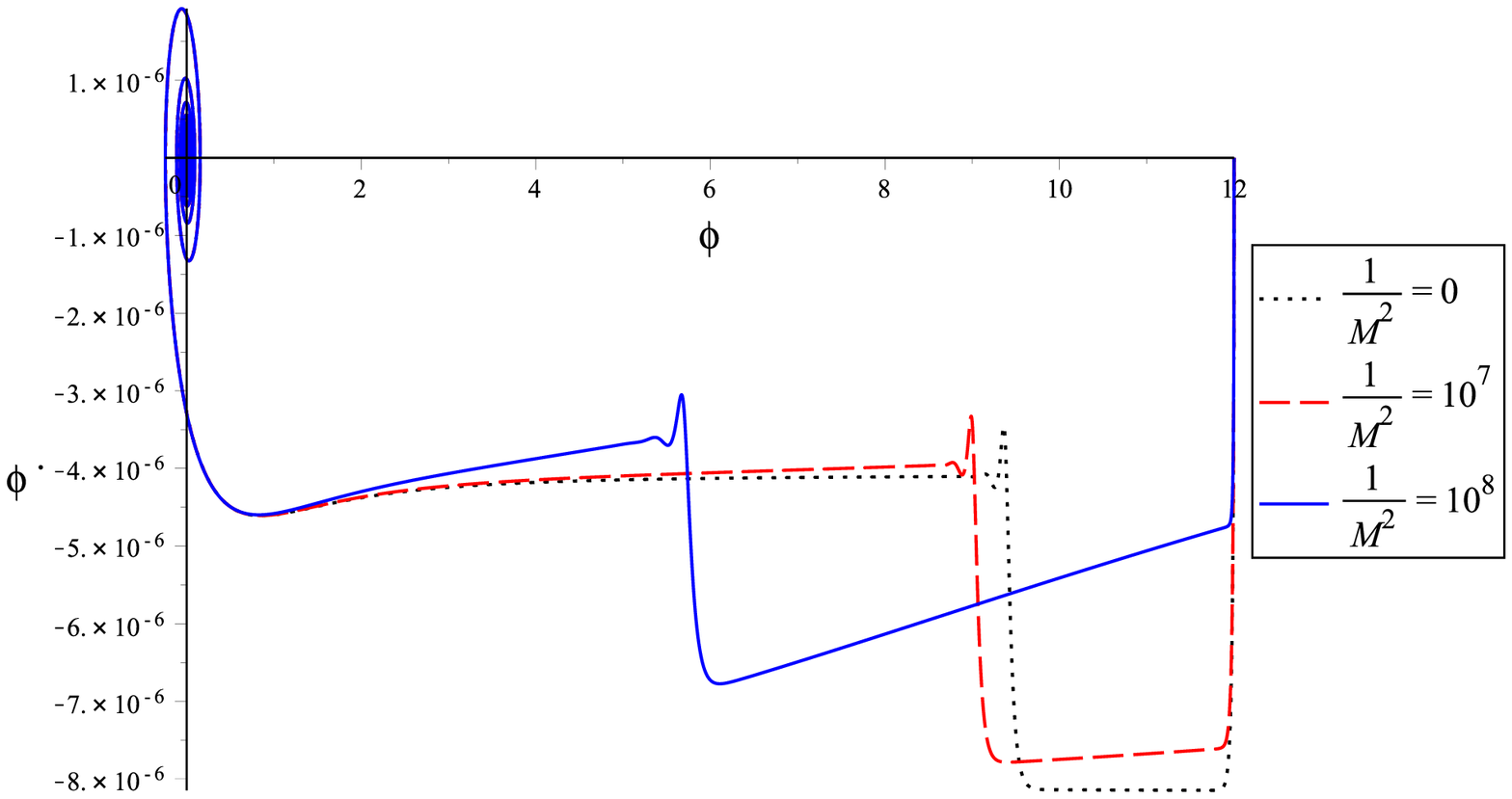}}
\end{center}
 \caption{The phase flow for $\phi$, in quadratic inflationary potential, for $c=2$ and different value of coupling constant $1/M^2$.}
\label{fig6}
\end{figure}
If $c<1$, then the anisotropy will rapidly decay. Hence, we get a condition for $c$, as
\begin{equation}\label{47}
{M_p^2\over2}{f^{'}(\varphi)V^{'}(\varphi)\over f(\varphi)V(\varphi)}\bigg(1+{V(\varphi)\over M^2M_p^2}\bigg)^{-1}>1.
\end{equation}
Therefore, any function pairs $f(\varphi)$ and $V(\varphi)$, which satisfies this condition in some range could generate vector hair during inflation.
We may ignore the effects of shear during inflation, then we can rewrite the equation of motion for scalar field (\ref{13}) during inflation as
\begin{equation}\label{48}
\Big(\ddot{\varphi}+3\dot{\alpha}\dot{\varphi}\Big)\Big(1+{3\dot{\alpha}^2\over M^2}\Big)+V'(\varphi)-2{f'(\varphi)\over f(\varphi)}\rho_v=0.
\end{equation}
Using equation (\ref{45}), we have
\begin{eqnarray}\label{49}
&&\Big(\ddot{\varphi}+3\dot{\alpha}\dot{\varphi}\Big)\Big(1+{3\dot{\alpha}^2\over M^2}\Big) \nonumber\\
&&+V'(\varphi)\Bigg[1-{2c\over\epsilon_v}\bigg[1-{V^{'}(\varphi)^2\over6M^2V(\varphi)}\Big(1+{V(\varphi)\over M^2M_p^2}\Big)^{-2}\bigg]\mathcal{R}\Bigg]=0,
\end{eqnarray}
where $\mathcal{R}$ and slow-roll parameter $\epsilon_v$ are defined as;
\begin{eqnarray}
\label{50}
\mathcal{R}&=&{\rho_v\over V(\varphi)}\bigg[1-{V'(\varphi)^2\over 6M^2V(\varphi)}\Big(1+{V(\varphi)\over M^2M_p^2}\Big)^{-2}\bigg]^{-1},\\
\label{51}
\epsilon_v&=&{M_p^2\over2}\bigg({V'(\varphi)\over V(\varphi)}\bigg)^2\bigg(1+{V(\varphi)\over M^2M_p^2}\bigg)^{-1}.
\end{eqnarray}
Also, using the slow-roll approximation, we can rewrite equation (\ref{48}) in the form of
\begin{equation}\label{52}
3\dot{\alpha}\dot{\varphi}\bigg(1+{3\dot{\alpha}^2\over M^2}\bigg)
+V'(\varphi)-P_A^2f(\varphi)^{-3}f'(\varphi)e^{-4\alpha-4\beta}=0.
\end{equation}
Now by substituting relation $\dot{\alpha}^2\approx {V(\varphi)/3M_p^2}$ and $f(\varphi)=\exp{(-2c\alpha )}$ into the relation (\ref{52}) we get
\begin{eqnarray}\label{53}
{\dot{\varphi}\over \dot{\alpha}}&\approx&
-M_p^2{V'(\varphi)\over V(\varphi)}\bigg(1+{V(\varphi)\over M^2 M_p^2}\bigg)^{-1}
+{2cP_A^2\over V'(\varphi)}\nonumber \\
&&\times\exp{\bigg[-4\alpha-4\beta-{4c\over M_p^2}\int{{V(\varphi)\over V'(\varphi)}\bigg(1+{V(\varphi)\over M^2 M_p^2}\bigg)d\varphi}\bigg]}.
\end{eqnarray}
If we ignore the evolution of $V(\varphi)$, $V'(\varphi)$ and $\beta$ with respect to $\alpha$, then we can solve this differential equation as
\begin{eqnarray}\label{54}
&&\exp{\bigg[-4\alpha-4\beta-{4c\over M_p^2}\int{{V(\varphi)\over V'(\varphi)}\bigg(1+{V(\varphi)\over M^2 M_p^2}\bigg)d\varphi}\bigg]}= \nonumber \\
&&\hspace*{1cm}{2c^2P_A^2V(\varphi)\over (c-1)M_p^2V'(\varphi)^2}\bigg(1+{V(\varphi)\over M^2M_p^2}\bigg)\bigg(1+\Omega e^{[4\beta-4(c-1)\alpha]}\bigg),
\end{eqnarray}
where $\Omega$ is defined by
\begin{equation}\label{55}
\Omega={A(c-1)M_p^2V'(\varphi)^2\over2c^2P_A^2 V(\varphi)},
\end{equation}
and $A$ is the constant of integration. Consequently by substituting the exponential part of equation (\ref{54}) back into the equation of motion (\ref{53}), we have
\begin{eqnarray}\label{56}
{\dot{\varphi}\over \dot{\alpha}}&\approx&
-M_p^2{V'(\varphi)\over V(\varphi)}\bigg(1+{V(\varphi)\over M^2M_p^2}\bigg)^{-1}\times\nonumber \\
&&\Bigg[1-\Big({c-1\over c}\Big)\bigg[1+\Omega\exp{\bigg(-4\alpha-4\beta-{4c\over M_p^2}\bigg)}\bigg]^{-1}\Bigg].
\end{eqnarray}
Initially, as $\alpha\rightarrow-\infty$, the quantity
\begin{equation}\label{56.1}
\bigg[1+\Omega\exp{\bigg(-4\alpha-4\beta-{4c\over M_p^2}\bigg)}\bigg]^{-1}\rightarrow0
\end{equation}
 so that conventional isotropic slow-roll inflationary phase in non-minimal derivative coupling model is encountered as described by equation (\ref{43}). On the other hand, in the limit $\alpha\rightarrow\infty$, the quantity
 \begin{equation}\label{56.2}
 \bigg[1+\Omega\exp{\bigg(-4\alpha-4\beta-{4c\over M_p^2}\bigg)}\bigg]^{-1}\rightarrow1.
 \end{equation}
So that
\begin{equation}\label{56.3}
{\dot{\varphi}\over \dot{\alpha}}\approx-{M_p^2\over c}{V'(\varphi)\over V(\varphi)}\bigg(1+{V(\varphi)\over M^2M_p^2}\bigg)^{-1}.
\end{equation}

\begin{figure}[t]
\begin{center}
  \scalebox{0.6}{\includegraphics{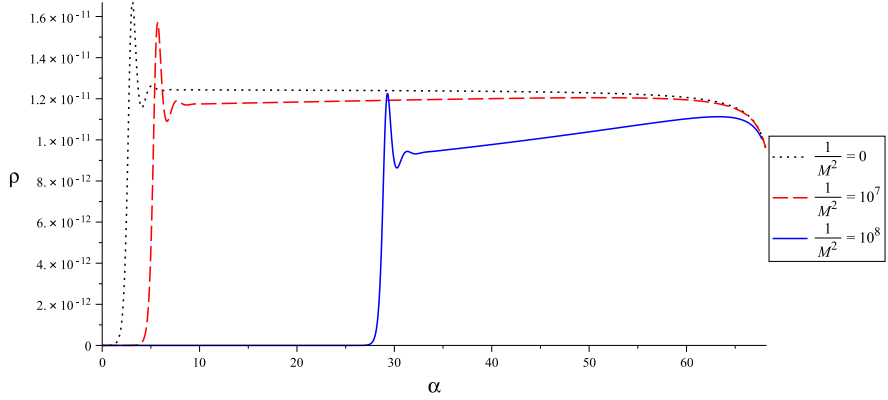}}
\end{center}
 \caption{The energy density of vector field ${\rho_v}$ in terms of the number of e-folds $\alpha$, in quadratic inflationary potential, for $c=2$ and different values of coupling constant $M$.}
\label{fig7}
\end{figure}
This relation as compared to (\ref{43}) is the second inflationary phase which is now accompanied with small effects of anisotropy.
We have seen that in the second phase of inflation (\ref{56.3}) is $1/c$ times reduced compared to (\ref{43}).
Obviously, there is a transition from  conventional isotropic slow-roll inflationary phase to what we refer to as the second inflationary phase.
Therefore, in the second slow-roll inflationary phase, equation (\ref{54}) becomes
\begin{eqnarray}\label{57}
\exp{\bigg[-4\alpha-4\beta-{4c\over M_p^2}\int{{V(\varphi)\over V'(\varphi)}\bigg(1+{V(\varphi)\over M^2 M_p^2}\bigg)d\varphi}\bigg]}= \nonumber \\
{(c-1)M_p^2V'(\varphi)^2\over 2c^2P_A^2 V(\varphi)}\bigg(1+{V(\varphi)\over M^2 M_p^2}\bigg)^{-1}.
\end{eqnarray}
Now using relation (\ref{14}), (\ref{45}) and (\ref{57}) the energy density of vector field $\rho_v$ as a function of inflaton potential can be expressed as,
\begin{eqnarray}\label{58}
\rho_v&=&{P_A^2\over2}\exp{\bigg[-4\alpha-4\beta-{4c\over M_p^2}\int{{V(\varphi)\over V'(\varphi)}\bigg(1+{V(\varphi)\over M^2 M_p^2}\bigg)d\varphi}\bigg]}
 \nonumber \\
&=&{M_p^2V'(\varphi)^2(c-1)\over 4c^2 V(\varphi)}\bigg(1+{V(\varphi)\over M^2 M_p^2}\bigg)^{-1}.
\end{eqnarray}
In the second inflationary phase we can see that the energy density of vector field is proportional to slow-roll parameters of the theory as follows
\begin{equation}\label{59}
\rho_v={c-1\over2c^2} \epsilon_v V(\varphi).
\end{equation}
Finally, after straightforward calculations, we can obtain a remarkable result
\begin{equation}\label{60}
{\Sigma\over H}\approx{c-1\over3c^2}\epsilon_v\bigg[1-{V'(\varphi)^2\over 6M^2V(\varphi)}\bigg(1+{V(\varphi)\over M^2M_p^2}\bigg)^{-2}\bigg]^{-1},
\end{equation}
corresponding to the anisotropic inflationary phase.
This relation show that the spatial shear is proportional to slow-roll parameter, always small and it will not away during inflation.
The ratio of energy density of vector field to that inflaton field is given by
\begin{equation}\label{60.1}
{\rho_v\over\rho_\varphi}={c-1\over2c^2}\epsilon_v.
\end{equation}
Specifically, from relation (\ref{56.3}), during anisotropic inflation the slow-roll parameter becomes
\begin{equation}\label{60.2}
\epsilon=-{\dot{H}\over H^2}=-{1\over2}{V'(\varphi)\over V(\varphi)}{\dot{\varphi}\over H}={1\over c}\epsilon_v.
\end{equation}

Let us consider two special cases of asymptotic values of the coupling constant $1/M^2$.
In the limit as $1/M^2\rightarrow0$, the ratio of shear to expansion rate from relation (\ref{60}) and slow-roll parameter from equation (\ref{51}) becomes
\begin{equation}\label{61}
{\Sigma\over H}\approx{c-1\over3c^2}\epsilon_v,   \;\;\;\;\;\;  \epsilon_v\approx {M_p^2\over2}\Big({V'(\varphi)\over V(\varphi)}\Big)^2.
\end{equation}
We have seen that these relations are in agreement with those the results of the anisotropic inflation in conventional minimal coupling model \cite{Soda,Soda1}.
In the other hand, high friction regime in non-minimal derivative coupling model defined as $(H^2/M^2)\gg1$, \cite{Germani1}. Thus, From equation (\ref{38}) we can obtain this condition for potential as follows;
\begin{equation}\label{62}
V(\varphi)\gg 3M^2M_p^2.
\end{equation}
Therefore the equation (\ref{60}) and slow-roll parameter (\ref{51}) in the high friction limits becomes;
\begin{equation}\label{63}
{\Sigma\over H}\approx{c-1\over3c^2}\epsilon_v, \;\;\;\;\;\;\;  \epsilon_v \approx {M_p^4M^2\over2V(\varphi)}\Big({V'(\varphi)\over V(\varphi)}\Big)^2.
\end{equation}
From relation (\ref{45}), the coupling function $f(\varphi)$ in this regime takes the simple form;
\begin{equation}\label{64}
f(\varphi)\approx \exp{\bigg({2c\over M_p^4M^2}\int\Big({V(\varphi)^2\over V'(\varphi)}\Big)d\varphi\bigg)}.
\end{equation}
These relations for power-law inflationary potential $V(\varphi)=\lambda \varphi^n$ becomes;
\begin{eqnarray}\label{65}
\epsilon_v &\approx& {M_p^4M^2n^2\over2\lambda\varphi^{n+2}}, \nonumber \\
\rho_v &\approx& {(c-1)M_p^4M^2n^2\over4c^2\varphi^2}, \nonumber \\
f(\varphi)&\approx& \exp{\bigg({2c\lambda\varphi^{(n+2)}\over n(n+2)M_p^4M^2}\bigg)}.
\end{eqnarray}
From slow-roll condition and high friction limit it is easy to obtain a condition
\begin{equation}\label{66}
{M^2 \over \lambda}\ll {n^n\over3\times6^{n\over2}}M_p^{(n-2)}.
\end{equation}
Therefore, for an abroad class of power-law potentials and coupling constant that satisfy in this condition there are anisotropic inflationary solution.

\section{Numerical analysis}

\begin{figure}[t]
\begin{center}
  \scalebox{0.6}{\includegraphics{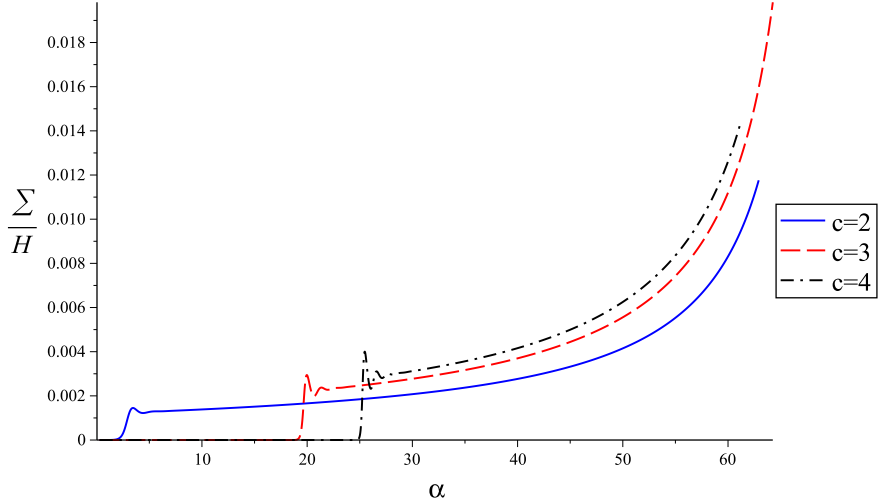}}
\end{center}
 \caption{Anisotropy $\Sigma/H$ in terms of the number of e-folds $\alpha$, in quadratic inflationary potential, for $c=2$ and different values of coupling constant $M$.}
\label{fig8}
\end{figure}

\begin{figure}[t]
\begin{center}
  \scalebox{0.6}{\includegraphics{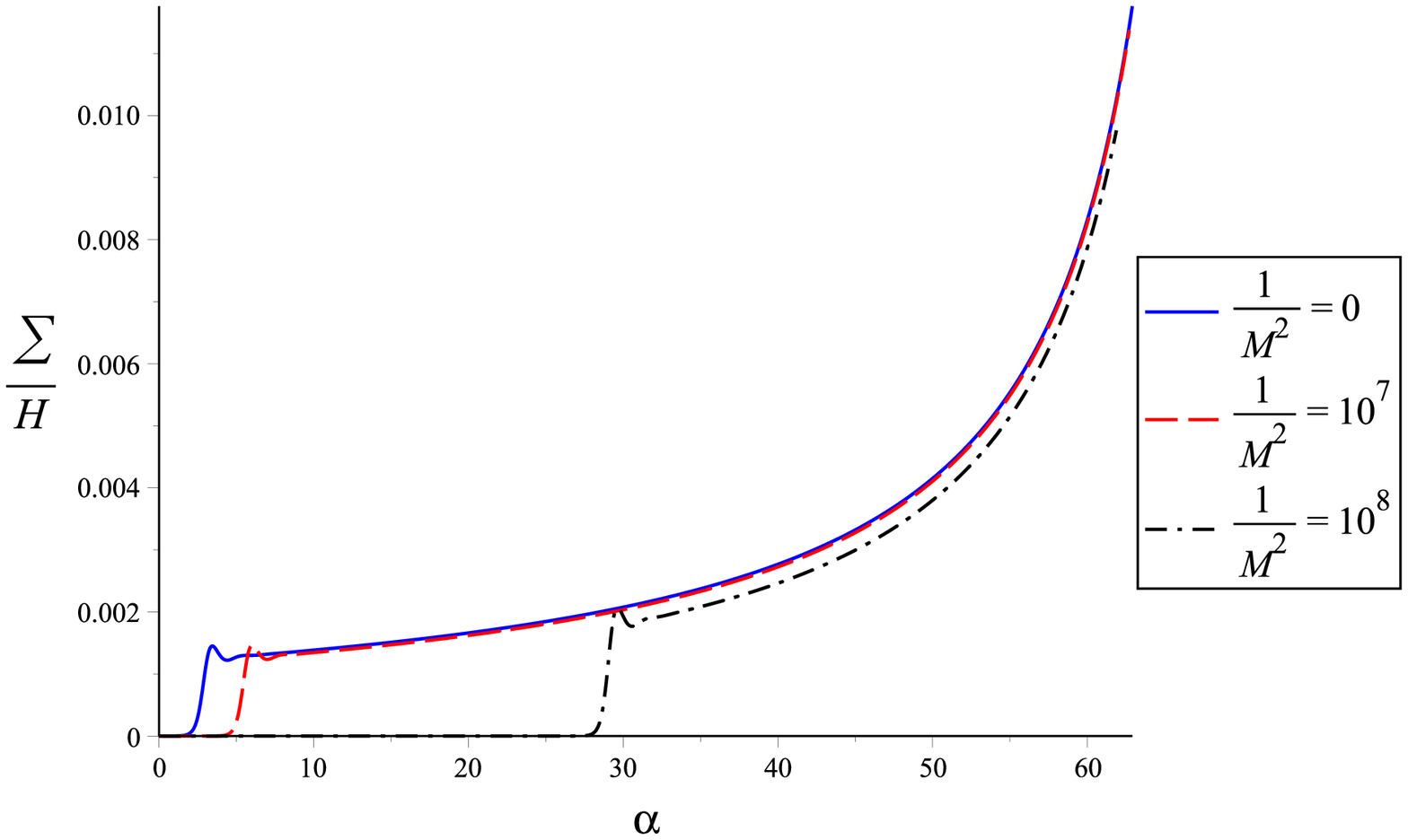}}
\end{center}
 \caption {Anisotropy $\Sigma/H$ in terms of the number of e-folds $\alpha$, in quadratic inflationary potential, for $c=2$ and different values of coupling constant $M$.}
\label{fig9}
\end{figure}

\begin{figure}[t]
\begin{center}
  \scalebox{0.6}{\includegraphics{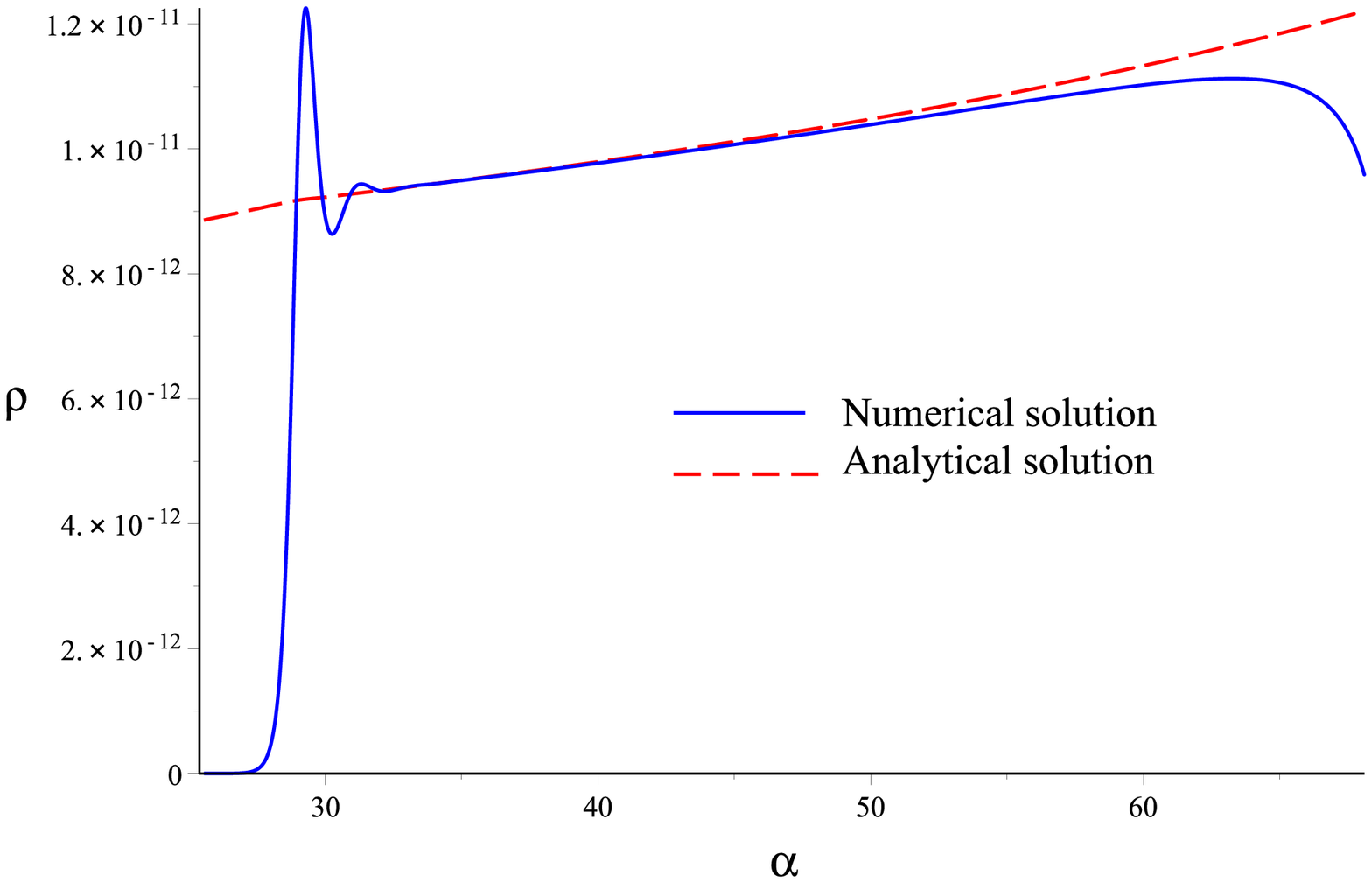}}
\end{center}
 \caption {The evolution of $\rho$ in terms of the number of e-folds $\alpha$, for quadratic inflationary potential, $1/M^2=10^8$ and $c=2$ is plotted. The solid blue curve is the full numerical result whereas the dashed red curve is the analytical solution.}
\label{fig10}
\end{figure}

\begin{figure}[t]
\begin{center}
  \scalebox{0.6}{\includegraphics{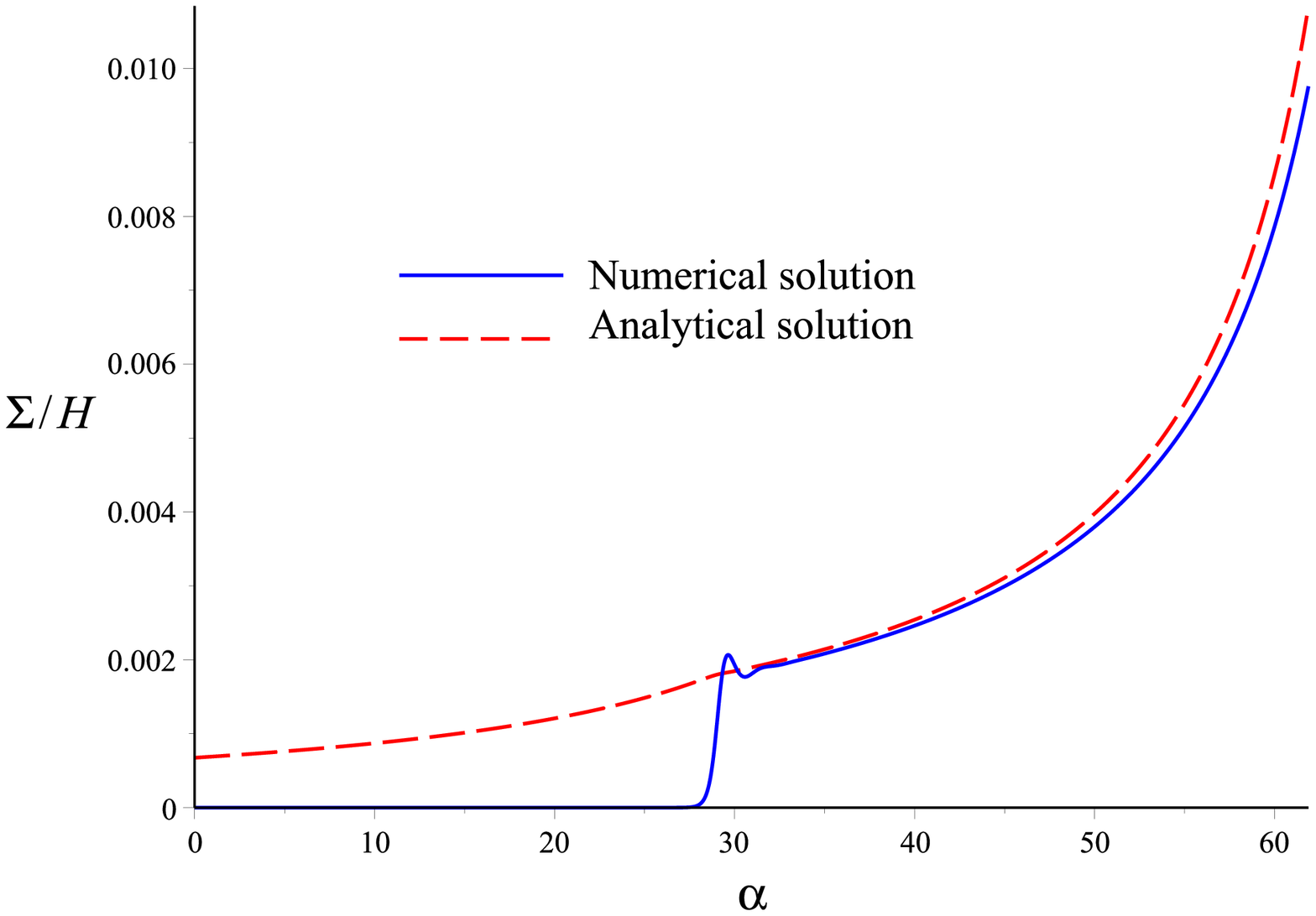}}
\end{center}
 \caption {The evolution of $\Sigma/H$ in terms of the number of e-folds $\alpha$, for quadratic inflationary potential, $1/M^2=10^8$ and $c=2$ is plotted. The solid blue curve is the full numerical result whereas the dashed red curve is the analytical solution.}
\label{fig11}
\end{figure}

In the previous section, we found the slow-roll inflation with the anisotropy in non-minimal derivative coupling model. From analytical investigation we have seen there are two phases of inflation, at the $\alpha\rightarrow-\infty$ where is named the conventional slow roll inflation and $\alpha\rightarrow+\infty$ where dubbed anisotropic slow-roll inflation or second phase.
Now, in this section we solve four field equations (\ref{10}-\ref{13}) numerically to compare this results whit a analytical finding.
Initial conditions where we have use in all numerical calculation in this paper are
\begin{equation}\label{66.1}
\varphi_0=12,\;\; \dot{\varphi}_0=0,\;\; \alpha_0=0,\;\; \beta_0=0,\;\; \dot{\beta}_0=0,
\end{equation}
and from relation (\ref{10}) we can calculate the boundary condition for $\dot{\alpha}_0$ as
\begin{equation}\label{67}
\dot{\alpha}_0=\sqrt{{(V(\varphi_0)+\rho_v(\varphi_0))/3M_p^2}}.
\end{equation}
Here $0$ subscript indicates the initial value. In this model we have two parameters, coupling constant $1/M^2$ and $c$ that in what follows we present
the results of the numerical analysis of the field equations (\ref{10}-\ref{13}) corresponding to different values of this parameters. In this plots we try to present this for minimal and non-minimal coupling for comparison.
All of this plots were generated with the quadratic potential as
\begin{equation}\label{68}
V(\varphi)={1\over 2}m^2\varphi^2,
\end{equation}
where $m$ is the mass of inflaton field, here we took this parameters $m=10^{-5}M_p$ in all of the plots.
We can obtain coupling function for this potential from relation (\ref{45}) as follows
\begin{equation}\label{68.1}
f(\varphi)=\exp{\bigg({c\over2M_p^2}\varphi^2+{cm^2\over 8M^2M_p^4}\varphi^4\bigg)}.
\end{equation}
Now, let us calculate the number of e-folds from the beginning of inflation to the end of inflation.
We have seen that there are two distinct phases of inflation: the first slow-roll phase (isotropic phase) and the second slow roll phase (anisotropic phase), where the ratio $\dot{\alpha}/\dot{\varphi}$ for them is given by relations (\ref{43}) and (\ref{56}) respectively.
The transition from isotropic slow-roll inflationary phase to anisotropic slow-roll inflationary phase occur at $t_{P.T}$.
Therefore, we can obtain the number of e-folds as follows
\begin{eqnarray}\label{69}
N&=&\int_{t_0}^{t_{end}}Hdt \nonumber \\
&=&\int_{t_0}^{t_{P.T}}\dot{\alpha}_{f}dt+\int_{t_{P.T}}^{t_{end}}\dot{\alpha}_{s}dt\nonumber \\
&=&\bigg(\alpha_f(t_{P.T})-\alpha_f(t_{0})\bigg)+\bigg(\alpha_s(t_{end})-\alpha_s(t_{P.T})\bigg),
\end{eqnarray}
where $t_{0}$ and $t_{end}$ are the cosmic time at the beginning and the end of inflation, respectively.
$\alpha_{f}$ and $\alpha_{s}$ are respectively the $\alpha$ function during the first slow-roll phase and the second slow-roll phase.
 From relation (\ref{43}) and (\ref{56}) we can  calculate the number of e-folds between beginning of inflation and the end of inflation for quadratic potential as
\begin{eqnarray}\label{70}
N&\approx&{\varphi_0^2\over4M_p^2}\bigg(1+{m^2\over 2M^2M_p^2}\varphi_0^2\bigg) \nonumber \\
&+&{(c-1)\over4}{\varphi_{P.T}^2\over M_p^2}\bigg(1+{m^2\over 2M^2M_p^2}\varphi_{P.T}^2\bigg).
\end{eqnarray}
Where $\varphi_0$ and $\varphi_{P.T}$ are the scalar field at the beginning of the inflation and the time of phase transition respectively.
We disregard the term $\alpha_s(t_{end})$ because, it is negligible in comparison with the other terms in equation (\ref{69}).

In Figure \ref{fig1} and Figure \ref{fig2} the growth of $\alpha$ as a cosmic time is presented.
We have seen that evolution of the number of e-folds $\alpha$ is not so sensitive to the change of parameters $c$ and $M$.

The slow-roll parameters for quadratic potential at the beginning of inflation, from relation (\ref{51}) becomes
\begin{equation}\label{71}
\epsilon_v={2M_p^2\over\varphi^2}\bigg(1+{m^2\over 2M^2M_p^2}\varphi^2\bigg)^{-1},
\end{equation}

Figure \ref{fig3} and Figure \ref{fig4} shows classical trajectory of the scalar field $\varphi$ versus the number of e-folds.
As we see there are a period of quasi-de Sitter inflation and the short period of rapid oscillation of scalar field for all of the parameters $c$ and $M$.
The enhancement of the parameter $c$ cause that the period of quasi-de Sitter inflation makes longer.
However two curves in Figure \ref{fig3} have the same initial condition, but the cure with the larger values of coupling constant $1/M^2$, the scalar field has slower evolution, due to the gravitationally enhancement friction mechanism.

Phase flow for $\varphi$ with different values of $M$ and $c$ is depicted in Figure \ref{fig5} and Figure \ref{fig6}.
Obviously we see that there are two slow-roll inflationary phases, similar to those an anisotropic inflation in minimal coupling model; isotropic phase and anisotropic phase.
It is clear that, each transition from isotropic to anisotropic phase occurred at the spacial value of scalar field $\varphi_{P.T}$ where this quantity associated with constant parameters of the theory.
Moreover we see that for different values of $M$ and $c$ there are attractor solutions.
Clearly, as the coupling constant $1/M^2$ or constant $c$ is decreased, the region associated with anisotropic slow-roll phase increases.

As a concrete example by substituting the values of the parameters $\varphi_0=12$, $m=10^{-5}$ ,$c=3$, $1/M^2=10^{4}$ and $\varphi_{P.T}=8$ from Figure  \ref{fig5}, into equation (\ref{70}), we have $N\approx 68$. This is in excellent agreement with the data of numerical results in Figure \ref{fig4}.

Figure \ref{fig7} the energy density of vector field $\rho_v$ with respect to the e-folds number with different values of $M$ is depicted.
We see that in the slow-roll isotropic inflationary phases the energy density of vector field is negligible.
At the beginning of the slow roll anisotropic inflation phase the energy density of vector field grows rapidly, then this quantity remains constant until the end of anisotropic inflation.
From relation (\ref{59}) the energy density of vector field for quadratic potential in the slow-roll anisotropic phase of inflation becomes

\begin{equation}\label{72}
\rho_v={(c-1)m^2M_p^2\over2c^2}\bigg(1+{m^2\varphi^2\over 2M^2M_p^2}\bigg)^{-1}.
\end{equation}

Evolution of the anisotropic shear ${\Sigma/H}$ with respect to the e-folding number for variable values of $c$ and $M$ are presented in Figure \ref{fig8} and Figure \ref{fig9}. As we see the enhancement of the parameters $1/M^2$ and $c$ leading to upward shifts the curves.
Moreover, in this cases the ratio of shear to expansion rete $\Sigma/H$ increases very sharply as the magnitude of  $1/M^2$ or $c$  increases.

We plot evolution of $\rho$ in terms of the number of e-folds $\alpha$, for quadratic inflationary potential, $1/M^2=10^8$ and $c=2$ in Figure \ref{fig10}. The solid blue curve is numerical result whereas the dashed red curve is our analytical solution equation (\ref{72}).
We emphasize that this analytical solution correspond to the anisotropic phase of inflation.
The agreement between our analytical solution valid for the second phase of inflation and numerical results is good.

From relation (\ref{60}) the ratio of shear to Hubble parameter for quadratic inflationary potential in the anisotropic phase is obtain as
\begin{equation}\label{73}
{\Sigma\over H}\approx{c-1\over3c^2}\epsilon_v\bigg[1-{m^2\over 3M^2}\Big(1+{m^2\over3M^2M_p^2}\varphi^2\Big)^{-2}\bigg]^{-1}.
\end{equation}
We have plotted our analytical solution for $\Sigma/H$ compared to the full numerical result, for quadratic inflationary potential, $1/M^2=10^8$ and $c=2$.
The solid blue curve is the full numerical result whereas the dashed red curve is the analytical solution equation (\ref{73}).
As can be seen they are in very good agrement in the second phase of inflation.

Finally, the end of anisotropic inflation is when $\epsilon\approx1$, then we can obtain scalar field at the end of inflation from relation (\ref{51}) as follows
\begin{equation}\label{74}
\phi_{end}\approx{MM_p\over m}\bigg(\sqrt{1+({4m^2\over cM^2})}-1\bigg)^{1\over2}.
\end{equation}

This relation show that, in this model the scalar field at the end of inflation is depends on parameter $M$ and $c$.

For our specific example we take the parameters $m = 10^{-5}$, $c=2$ and $1/M^2=10^8$. It has the inflaton field at the end of inflation $\phi_{end}\approx M_p$ using equation (\ref{74}), the anisotropy $\Sigma/H\approx0.083$ using equation (\ref{73}). These results are in very good agreement with each other and with the value obtained from numerical data in all of the plots.

\section{Conclusion}

In the present work, we have studied anisotropic inflation model, in presence of non-minimal derivative coupling term and a massless $U(1)$ gauge field whose kinetic part is coupled to the scalar field.
The power-law solutions for anisotropic inflation in the context of non-minimal derivative coupling models in the high friction regime are obtained.

In this scenario, we have showed that power-law anisotropic inflationary solutions can be constructed when both the inflaton potential and gauge kinetic coupling function are the power-law depends on the inflaton field.

In presence of non-minimal derivative coupling, we have derived gauge coupling function as an exponential type depends on the inflaton field, and a general relation for anisotropy $\Sigma/H$ which show that anisotropy is proportional to the slow-roll parameter of the theory.
Our investigation demonstrate that anisotropic hair can survive during inflation in non-minimal derivative coupling model.

We showed both numerically and analytically that there are two phases of inflation, similar to those an anisotropic inflation in minimal coupling model, isotropic and anisotropic phase.
The transition from isotropic phase to anisotropic phase occurred in $\varphi_{P.T}$ where this quantity depends on non-minimal coupling constant $1/M^2$ and constant parameter $c$.
So that, increase of these parameters increases the number of e-folds during isotropic phase of inflation.

We carried out the numerical calculation for the quadratic inflationary potential and we saw that there are attractor solutions for different values of parameters $M$ and $c$.
In summary our numerical and analytical investigation shows that for wide range of initial condition and coupling constant $M$, anisotropy grows exponentially at the end of isotropic phase of inflation.

\end{document}